# Coupling angle resolved photoemission data and quasiparticle structure in antiferromagnetic insulators $Sr_2CuO_2Cl_2$ and $Ca_2CuO_2Cl_2$

V.Gavrichkov, S.Ovchinnikov, A.Borisov

660036, Kirensky Institute of Physics, Krasnoyarsk, Siberian Branch RAS, Russia

We have analyzed the quasiparticle dispersion and *ARPES*-spectral density for $Sr_2CuO_2Cl_2$ and $Ca_2CuO_2Cl_2$ antiferromagnetic insulators along basic symmetric directions of the Brillouin zone (*BZ*) in a framework of an extended tight binding method (*ETBM*) with explicit account for strong electron correlations. The quasiparticle dispersion is in a good agreement with *ARPES*- data. At the top of valence band we found a narrow impurity-like virtual level with the spectral weight proportional to the concentration of spin fluctuations. A pseudogap between the virtual level and the top of the valence band has dispersion similar to "remnant Fermi surface" in $Ca_2CuO_2Cl_2$ and to a pseudogap in the underdoped *Bi2212* samples. A calculated parity of the polarized *ARPES*-spectra in $\Gamma, M, X$ - points in the *AFM* phase is even with regard to relative magnitudes of the partial contributions by different orbitals to the total *ARPES*- spectral density. Conditions of an observability for the different partial contributions in the polarized *ARPES*- experiments are determined also.

## I. Introduction

An investigation of the antiferromagnetic insulators $Sr_2CuO_2Cl_2$ and $Ca_2CuO_2Cl_2$ by the angle-resolved photoemission spectroscopy (*ARPES*) is one of ways to study of an evolution of the electronic structure of *HTSC*- materials with doping. Records of *ARPES*- spectra along the basic symmetric $\vec{k}$ - directions for different polarizations of a radiation allow to classify the valence band states in agreement with their symmetry properties. Really, in a structure of $Sr_2CuO_2Cl_2$ and $Ca_2CuO_2Cl_2$ it is possible to select three groups of the valence states. The first group are an invariant concerning a reflection in a plane of a photoemission. The second and third groups are accordingly even and odd concerning the same reflection. According to selection rules [1] in *ARPES*- spectra with a parallel vector of a polarization with respect to a photoemission plane the even group, and in spectra with perpendicular polarization only the odd group of states will be observed only. In the perpendicular geometry a vector of a polarization is always parallel to $CuO_2$ - plane. Therefore, a contribution from the valence states linking the planes in a uniform three-dimensional structure will be seen only in spectra with the parallel geometry. Now, we sum the basic results of *ARPES*- investigations touching in the present work, as following:

1. Analysis of the occupation $n(\vec{k})$ [2], obtained from *ARPES*- spectra of the oxychlorides $Sr_2CuO_2Cl_2$ and $Ca_2CuO_2Cl_2$, shows a singularity in $n(\vec{k})$ at crossing of $\vec{k}$ - contour, close to a contour of Fermi- surface predicted earlier by the *LDA*-calculation. A dispersion of the quasiparticle peak along $\vec{k}$ - contour of "remnant Fermi surface" is close to $|cos\ (k_xa) - cos(k_ya)|$ d-like dispersion. As the last is similar to a dispersion of the pseudogap in the underdoped $Bi_2Sr_2CaCu_2O_{8+\delta}(Dy)$ (*Bi2212*) and superconductivity gap in the optimal doped *Bi2212*- materials, there is a clear connection between all three energy gaps.

2. On the other hand, $|cos\ (k_xa) - cos(k_ya)|$ - dependence has a linear character near $\overline{M} = (\pi/2; \pi/2)$ - point. The same linear character is present in the spinon dispersion: $\sim J\sqrt{\cos^2(k_xa) - \cos^2(k_ya)}$ [3]. However, the experimental dispersion has more the quadratic behavior near $\overline{M}$, than the linear one [4].

3. Despite of a good reproducing of the observed dispersion in a framework of the *t-t'-t''-J* model, in its application to the real HTSC- materials there is a problem. In the *t-t'-t''-J* - model a dispersion along $\Gamma = (0,0) \leftrightarrow M(\pi,\pi)$ and $X = (0,\pi) \leftrightarrow Y = (\pi,0)$ also is stipulated by the parameters *J* and *t'* - accordingly. Therefore, the observed similarity in the dispersions along these directions superimposes improbably rigid restrictions on the parameters of the model. Available explanations of a universality of this similarity depended on a approximation used in a calculation. The self-consistent Born approximation to the *t-t'-t''-J* model is used in [5].

4. In the experiments with a polarized radiation the even parity of *ARPES*-spectra in the undoped oxichlorides in the $\Gamma, M, X, Y$ - symmetric points was found [6]. The parity varies with a doping. In a framework of the *t-t'-t''-J* model it is impossible to interpret a parity. As a consequence, the theoretical works touching analysis of the polarized *ARPES*-spectra, are based on an approximation of the local density approximation (*LDA*) [7]. The results obtained in frameworks of *LDA*, allow analyzing parity. For the dispersion, however, there isn't an analogy with successful results of the *t-t'-t''-J* model. In particular, *LDA* did not reproduce the *ZR*- quasiparticle peak on a top of the valence band [8] and failed to obtain the insulator ground state of undoped $CuO_2$- plane.

In the present work we analyze the spectral density for the $Sr_2CuO_2Cl_2$ and $Ca_2CuO_2Cl_2$- oxichlorides, recorded at the different polarizations of a radiation. Results obtained here with explicit treatment of strong electron correlations within a framework of an extended tight binding method (*ETBM*) [9], provides a natural generalization of results of *t-t'-t''-J* - model and admit a clear physical explanation.

The Sec.II contains a brief description of a derivation of the *ETBM*- Hamiltonian and basic formulas concerning with the dispersion and spectral density.

In the Sec. III the results of a computer simulation of the dispersion, and also amplitude of the quasiparticle peak in *ARPES*- spectral density along $\Gamma \leftrightarrow M \leftrightarrow X \leftrightarrow \Gamma$ and $X \leftrightarrow Y$ directions in the paramagnetic phase (*PM*) and antiferromagnetic phase (*AFM*) are discussed. Moreover, the partial contributions to the spectral density from the different orbitals are calculated. It is important from the point of view of an identification of *ARPES*-spectra, as they can have the different parity and cross-section of scattering of an incident radiation. A nature of the energy gap and dispersion along $\vec{k}$ - contour of the "remnant Fermi surface" is discussed also.

In the Sec.IV it is carried out a symmetry analysis of the partial contributions in $\Gamma, M, X, Y$ - symmetry points. As a consequence, aspects of a polarization are indicated at which it is possible to observe the different contributions.

The Sec.V considers an effect of spin fluctuations on the band structure of the oxichlorides. The summary Sec.VI contains a brief enumeration of our conclusions.

## II. Tight binding method for strongly correlated electron system

In this part the brief description of the *ETBM* will be given, where a unit cell will be $CuO_4Cl_2$ ($CuO_6$) cluster, and a problem of a nonorthogonality of the molecular orbitals of adjacent clusters will be solved by an explicit fashion – by constructing of relevant Wannier functions on a five-orbitals initial basis of the atomic states. In a new symmetric basis an one-cells part of the total Hamiltonian is factorized, allowing to classify all possible effective quasiparticle excitations in $CuO_2$- plane according to a symmetry. A next step is a subsequent exact diagonalization of the Hamiltonian of a unit cell and transfer to a representation of Hubbard operators. Then we will consider an intercell part of a total Hamiltonian in a framework of the Hubbard 1- approximation. An initial Hamiltonian of a multiband p-d model can be wrote as:

$$H = H_d + H_p + H_{pd} + H_{pp}, \quad H_d = \sum_r H_d(r), \quad (1)$$

$$H_d(r) = \sum_{\lambda\sigma}\left[(\varepsilon_\lambda - \mu)d^+_{\lambda r\sigma}d_{\lambda r\sigma} + \tfrac{1}{2}U_\lambda n^\sigma_{\lambda r}n^{-\sigma}_{\lambda r} + \sum_{\lambda'\sigma'}\left(-J_d d^+_{\lambda r\sigma}d_{\lambda r\sigma'}d^+_{\lambda' r\sigma'}d_{\lambda' r\sigma} + \sum_{r'}V_{\lambda\lambda'}n^\sigma_{\lambda r}n^{\sigma'}_{\lambda' r'}\right)\right]$$

$$H_p = \sum_i H_p(i), \quad H_p(i) = \sum_{\alpha\sigma}\left[(\varepsilon_\alpha - \mu)p^+_{\alpha i\sigma}p_{\alpha i\sigma} + \tfrac{1}{2}U_\alpha n^\sigma_{\alpha i}n^{-\sigma}_{\alpha i} + \sum_{\alpha' i'\sigma'}V_{\alpha\alpha'}n^\sigma_{\alpha i}n^{\sigma'}_{\alpha' i'}\right]$$

$$H_{pd} = \sum_{<i,r>} H_{pd}(i,r), \quad H_{pd}(i,r) = \sum_{\alpha\lambda\sigma\sigma'}\left(t_{\lambda\alpha}p^+_{\alpha i\sigma}d_{r\lambda\sigma} + V_{\alpha\lambda}n^\sigma_{\alpha i}n^{\sigma'}_{\lambda r}\right)$$

$$H_{pp} = \sum_{<i,j>}\sum_{\alpha\beta\sigma}\left(t_{\alpha\beta}p^+_{\alpha i\sigma}p_{\beta j\sigma} + \text{э.с.}\right),$$

where $n^\sigma_{\lambda r} = d^+_{\lambda r\sigma}d_{\lambda r\sigma}; n^\sigma_{\alpha i} = p^+_{\alpha i\sigma}p_{\alpha i\sigma}$. r and i – run through positions of copper $d_{x^2-y^2} \equiv d_x, d_{3z^2-r^2} \equiv d_z$ - and oxygen $p_x, p_y, p_z$ - atomic orbitals.

$\varepsilon_\lambda = \varepsilon_{d_x}(\lambda = d_x)$ ; $\varepsilon_{d_z}(\lambda = d_z)$ - and $\varepsilon_\alpha = \varepsilon_p(\alpha = p_x, p_y)$ ; $\varepsilon_{p_z}(\alpha = p_z)$ - energies of the corresponding atomic orbitals; $t_{\lambda\alpha} = t_{pd}(\lambda = d_x; \alpha = p_x, p_y)$, $t_{pd}/\sqrt{3}(\lambda = d_z, \alpha = p_x, p_y)$ - matrix elements of a hopping; $U_\lambda = U_d(\lambda = d_x, d_z)$ and $U_\alpha = U_p(\alpha = p_x, p_y, p_z)$ - intraatomic Coulomb interactions; $V_{\alpha\lambda} = V_{pd}(\alpha = p_x, p_y; \lambda = d_x, d_z), V'_{pd}(\alpha = p_z; \lambda = d_x, d_z)$ - Coulomb repulsion between copper and oxygen. The stroke relates to the apical positions.

Each atom of the in-plane oxygen belongs to two cells simultaneously, so there is the problem of a nonorthogonality. It is actually more convenient to transform from the in-plane oxygen orbitals in $p_x^\sigma O, p_y^\sigma O, p_z^\sigma Cl, d_x, d_z Cu$ - five-orbital basis to the Wannier orbitals with the appropriate symmetry [10].

$$\begin{pmatrix} b_{k\sigma} \\ a_{k\sigma} \end{pmatrix} = \hat{S} \begin{pmatrix} p_{xk\sigma} \\ p_{yk\sigma} \end{pmatrix} = \begin{pmatrix} is_x/\mu_k & is_y/\mu_k \\ is_y/\mu_k & -is_x/\mu_k \end{pmatrix} \begin{pmatrix} p_{xk\sigma} \\ p_{yk\sigma} \end{pmatrix}, \text{ where } \mu_k^2 = s_x^2 + s_y^2, |\hat{S}| = 1; \qquad (2)$$

Now, the initial Hamiltonian reduces to the Hamiltonian of a square quasi-planar array of cells, each of which contains five orbitals: $d_x, b, d_z, a, p_z$. The total Hamiltonian of the model may be rewritten in the form $H = H_c + H_{cc}$, where $H_c$ is the Hamiltonian for noninteracting cells and $H_{cc}$ - a cell-cell interaction.

$$H = H_c + H_{cc}, \quad H_c = \sum_{f\sigma} H_{f\sigma}; \quad H_{f\sigma} = h^{(a)} + h^{(b)} + h^{(ab)}; \qquad (3)$$

$$h^{(b)} = \left(\varepsilon_b n_b^\sigma + \varepsilon_{d_x} n_{d_x}^\sigma\right) + \tfrac{1}{2}U_d n_{d_x}^\sigma n_{d_x}^{-\sigma} + \tfrac{1}{2}U_b n_b^\sigma n_b^{-\sigma} + \sum_{\sigma'} V_{pd} n_{d_x}^\sigma n_b^{\sigma'} - \tau_b\left(d_{x\sigma}^+ b_\sigma + h.c.\right);$$

$$h^{(a)} = \left(\varepsilon_a n_a^\sigma + \varepsilon_{d_z} n_{d_z}^\sigma + \varepsilon_{p_z} n_{p_z}^\sigma\right) + \tfrac{1}{2}U_d n_{d_z}^\sigma n_{d_z}^{-\sigma} + \tfrac{1}{2}U_a n_a^\sigma n_a^{-\sigma} + \tfrac{1}{2}U'_p n_{p_z}^\sigma n_{p_z}^{-\sigma} + \sum_{\sigma'}\left(V'_{pd} n_{d_z}^\sigma n_{p_z}^{\sigma'} + V_{pd} n_{d_z}^\sigma n_a^{\sigma'}\right) +$$

$$+ \tau_a\left(d_{z\sigma}^+ a_\sigma + h.c\right) - \tau'_{pd}\left(d_{z\sigma}^+ p_{z\sigma} + h.c.\right) - t'_{pp}\left(a_\sigma^+ p_{z\sigma} + h.c\right);$$

$$h^{(ab)} = \sum_{\sigma'} U_d n_{d_x}^\sigma n_{d_z}^{\sigma'} + U_{ab} n_a^\sigma n_b^{\sigma'} + V_{pd} n_b^\sigma n_a^{\sigma'} + V_{pd} n_b^\sigma n_{d_z}^{\sigma'} + V'_{pd} n_{d_x}^\sigma n_{p_z}^{\sigma'};$$

$$H_{cc} = \sum_{(i\neq j)}\sum_\sigma \left(h_{hop}^{(b)} + h_{hop}^{(a)} + h_{hop}^{(ab)}\right);$$

$$h_{hop}^{(b)} = -2t_{pd}\mu_{ij}\left(d_{xi\sigma}^+ b_{j\sigma} + b_{i\sigma}^+ d_{xj\sigma}\right) - 2t_{pp}\nu_{ij} b_{i\sigma}^+ b_{j\sigma};$$

$$h_{hop}^{(a)} = \frac{2t_{pd}}{\sqrt{3}}\lambda_{ij}\left(d_{zi\sigma}^+ a_{j\sigma} + h.c.\right) + 2t_{pp}\nu_{ij} a_{i\sigma}^+ a_{j\sigma} - 2t'_{pp}\lambda_{ij}\left(p_{zi\sigma}^+ a_{j\sigma} + h.c.\right);$$

$$h_{hop}^{(ab)} = \frac{2t_{pd}}{\sqrt{3}}\xi_{ij}\left(d_{zi\sigma}^+ b_{j\sigma} + h.c.\right) + 2t_{pp}\chi_{ij}\left(a_{i\sigma}^+ b_{j\sigma} + h.c.\right) - 2t'_{pp}\xi_{ij}\left(p_{zi\sigma}^+ b_{j\sigma} + h.c.\right);$$

where $\varepsilon_b = \varepsilon_p - 2t_{pp}\nu_{00}$; $\varepsilon_a = \varepsilon_p + 2t_{pp}\nu_{00}$; $\tau_b = 2t_{pd}\mu_{00}$; $\tau_a = 2t_{pd}\lambda_{00}/\sqrt{3}$;

$\tau'_{pd} = 2t'_{pd}/\sqrt{3}$; $\tau'_{pp} = 2t'_{pp}\lambda_{00}$;

The coefficients: $\mu_{ij}, \nu_{ij}, \lambda_{ij}$ concern with a hybridization of states of the same symmetry and depend only on a distance between i- and j-cells. The coefficients $\xi_{ij}$ and $\chi_{ij}$ also concern with a hybridization of states, belonging to the different $a_1$- and $b_1$ - representations, and change a sign at reflection along the one of *x*- or *y*-axes. These coefficients were calculated, for example, in the works [9,10].

An exact diagonalization of $H_c$ is carried out separately in different sectors of a Hilbert space in according to the $n_h$- number of holes in a cell. In the vacuum sector ($n_h=0$) we have one state $|0\rangle = |d^{10}p^6\rangle$. In the sector $n_h=1$ there are the spin doublets: $|\tilde{a}_{1\sigma}\rangle$ and $|\tilde{b}_{1\sigma}\rangle$. $|\tilde{b}_{1\sigma}\rangle$ - ground state has $b_{1g}$ –symmetry and is a consequence of a hybridization of $d_x$ state with $b$ - oxygen orbitals. $|\tilde{a}_{1\sigma}\rangle$ - nearest excited state has $a_{1g}$ -symmetry arises as a result of mixing of $d_z$- copper state with the $a$ - states of O and $p_z$ - states of Cl. In the two-hole sector $n_h=2$ there is a set of the 36 - spin singlets and 28 - triplets of a different orbital symmetry. The lowest $|\tilde{A}_1\rangle$ - and nearest excited $|\tilde{B}_{1M}\rangle$ - states have the $^1A_{1g}$ - and $^3B_{1g}$ - symmetry. The ZR-singlet gives essential, but not the only one contribution to $|\tilde{A}_1\rangle$. The energy of the $|\tilde{B}_{1M}\rangle$ - triplet state is close enough to the energy of $|\tilde{A}_1\rangle$ - singlet, and even a crossover between them is possible (see Fig.1a,b,c). We find two ways for a stabilization of $|\tilde{B}_{1M}\rangle$ - state as a ground state in the two-hole sector:

- A decrease of the crystal field parameter $\Delta_d = \varepsilon_{d_z} - \varepsilon_{d_x}$ results in an increase of the $Cu^{3+}$ fraction in $|\tilde{B}_{1M}\rangle$. For this way, the fraction of $d_x d_z$ - Hund configuration, through which the population of $d_z$ - orbitals could be observed, increases.

- As energy of *p* - orbitals of Cl (O) at the apical position and the distance to the apical position decrease, together with the

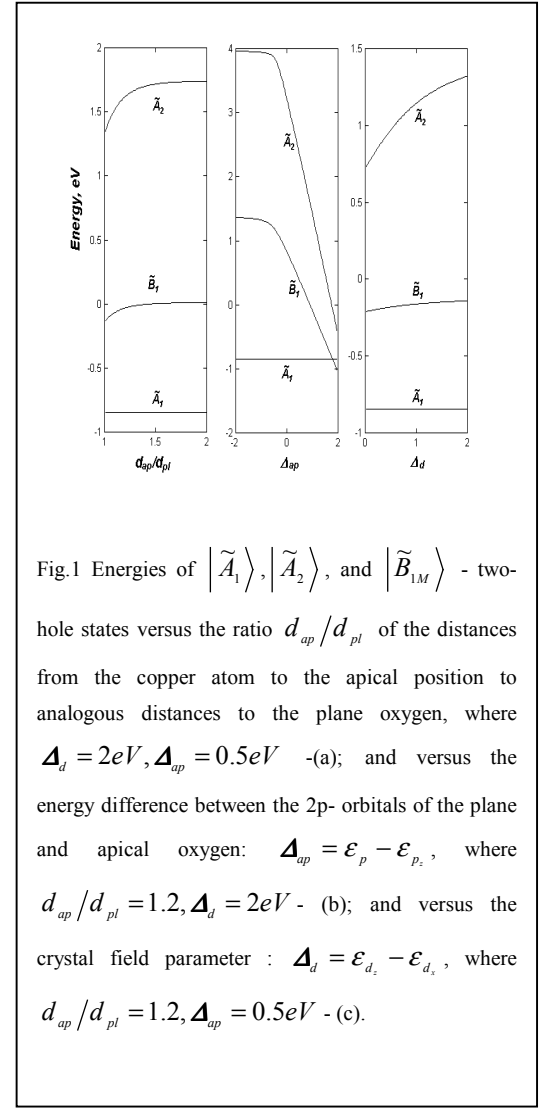

Fig.1 Energies of $|\tilde{A}_1\rangle, |\tilde{A}_2\rangle$, and $|\tilde{B}_{1M}\rangle$ - two-hole states versus the ratio $d_{ap}/d_{pl}$ of the distances from the copper atom to the apical position to analogous distances to the plane oxygen, where $\Delta_d = 2eV, \Delta_{ap} = 0.5eV$ -(a); and versus the energy difference between the 2p- orbitals of the plane and apical oxygen: $\Delta_{ap} = \varepsilon_p - \varepsilon_{p_z}$, where $d_{ap}/d_{pl} = 1.2, \Delta_d = 2eV$ - (b); and versus the crystal field parameter : $\Delta_d = \varepsilon_{d_z} - \varepsilon_{d_x}$, where $d_{ap}/d_{pl} = 1.2, \Delta_{ap} = 0.5eV$ - (c).

tendency for a crossover, there is also an appreciable increase of fraction of $d_x p_z$-symmetrized configuration in $|\widetilde{B}_{1M}\rangle$.

According to our estimates, the energy gap between the triplet and singlet states: $\Delta\varepsilon_2 \sim 0,5 eV$ for the realistic model parameters [9]. A presence of two states $|\widetilde{A}_1\rangle$ and $|\widetilde{B}_{1M}\rangle$, competing in energy leads to a necessity of their simultaneous consideration in our model and a impossibility of a further reduction to the effective one-band Hubbard model or $t$-$t'$-$t''$-$J$ – model. The singlet-triplet proximity influences very much a dispersion of a top of the valence band. In the new eigenstate basis the one-cell path of the total Hamiltonian of $CuO_2$ plane has a form:

$$H_c = \sum_{pf_G\sigma}(\varepsilon_{1pG} - \mu)X^{pp}_{f_G\sigma} + \sum_{qf_G\sigma}(\varepsilon_{2qG} - 2\mu)X^{qq}_{f_G\sigma} \text{ where } f_G = \begin{cases} f_A, f \in A \\ f_B, f \in B \end{cases}; \quad (4)$$

Here $p$ and $q$ number the one-hole and two-hole terms of a cell. $X^{pq}_f = |p\rangle\langle q|$ - Hubbard operators, constructed on the eigenstates. The levels of sublattices are split by a staggered field of the antiferromagnetic state: $\varepsilon_{1pA} = \varepsilon_{1p} - \sigma h$, $\varepsilon_{1pB} = \varepsilon_{1p} + \sigma h$. The magnitude $h \sim J\langle S_z\rangle$, where $J$- an effective exchange interaction of the nearest neighbors. $h$ decreases with doping, being converted in zero in the $PM$. We are limited by a nonself-consistent calculation, in which a magnetic state is considered as given ($PM$ or $AFM$ $T = 0K$). In the new eigenstate basis the one-electron operators are:

$$c_{f\lambda\sigma} = \sum_m \gamma_{\lambda\sigma}(m) X^m_{f\sigma} \quad (5)$$

where $c_{\lambda f\sigma} = d_{xf\sigma}, d_{zf\sigma}, a_{f\sigma}, b_{f\sigma}, p_{zf\sigma}$ and $m$- number a "rooted vector" $\vec{\alpha}_m(pq)$ [11]. The matrix elements $\gamma_{\lambda\sigma}(m)$ (m=0,1.. 31), corresponding to these "rooted vectors", are calculated straightforwardly using results of the exact diagonalization of the $H_c$. We take into account only two lowest terms ($\widetilde{b}_{1\sigma}$ and $\widetilde{a}_{1\sigma}$) in the one-hole and $\widetilde{A}_1$ - and $\widetilde{B}_{1M}$ - two-hole sectors. Therefore, $|p\rangle = |\widetilde{a}_{1\sigma}\rangle, |\widetilde{b}_{1\sigma}\rangle$ and $|q\rangle = |\widetilde{A}_1\rangle, |\widetilde{B}_{1M}\rangle$ in (5). Remaining terms stay much above in energy and we do not take them in the calculation of a spectra of the low-energy excitations.

A dispersion of the valence band was obtained with a help of the equations of a motion for the Green function

$$G^{\lambda\lambda'}_{k\sigma} = \langle\langle c_{k\lambda\sigma}|c^+_{k\lambda'\sigma}\rangle\rangle_E = \sum_{mn}\gamma_{\lambda\sigma}(m)\gamma^+_{\lambda'\sigma}(n)D^{mn}_{k\sigma}, \quad (6)$$

where: $\hat{D}_{k\sigma} = \begin{pmatrix} \hat{D}_{k\sigma}(AA) & \hat{D}_{k\sigma}(AB) \\ \hat{D}_{k\sigma}(BA) & \hat{D}_{k\sigma}(BB) \end{pmatrix}$, $D^{mn}_{k\sigma}(AB) = \langle\langle X^m_{k\sigma}|Y^n_{k\sigma}\rangle\rangle_E$

Eventually, in a framework of the Hubbard 1 approximation, the dispersion equation is determined by the equation:

$$\left\| (E - \Omega^G_m)\delta_{mn} / F^G_\sigma(m) - 2\sum_{\lambda\lambda'} \gamma^*_{\lambda\sigma}(m) T^{PG}_{\lambda\lambda'}(\vec{k}) \gamma_{\lambda'\sigma}(n) \right\| = 0 \qquad (7)$$

where $T^{AA}_{\lambda\lambda'}(k) = T^{BB}_{\lambda\lambda'}(k) = \frac{2}{N}\sum_{R_1} T^{AA}_{\lambda\lambda'}(\vec{R}_1) e^{ikR_1}$, $T^{AB}_{\lambda\lambda'}(k) = T^{BA}_{\lambda\lambda'}(k) = \frac{2}{N}\sum_{R_2} T^{AB}_{\lambda\lambda'}(\vec{R}_2) e^{ikR_2}$. In $d_x, d_z, b, a, p_z$ - basis the hopping matrix looks like:

$$T_{\lambda\lambda'}(\vec{R}) = \begin{pmatrix} 0 & 0 & -2t_{pd}\mu_{ij} & 0 & 0 \\ 0 & 0 & 2t_{pd}\xi_{ij}/\sqrt{3} & 2t_{pd}\lambda_{ij}/\sqrt{3} & 0 \\ -2t_{pd}\mu_{ij} & 2t_{pd}\xi_{ij}/\sqrt{3} & -2t_{pp}\nu_{ij} & 2t_{pp}\chi_{ij} & -2t'_{pp}\xi_{ij} \\ 0 & 2t_{pd}\lambda_{ij}/\sqrt{3} & 2t_{pp}\chi_{ij} & 2t_{pp}\nu_{ij} & -2t'_{pp}\lambda_{ij} \\ 0 & 0 & -2t'_{pp}\xi_{ij} & -2t'_{pp}\lambda_{ij} & 0 \end{pmatrix} \qquad (8)$$

The equation (7) is similar to the usual one-electron equation in a tight binding method, differing from it by two moments. At first, quasiparticle energies are defined by $\Omega^G_m = \varepsilon_{2qG} - \varepsilon_{1pG}$ - resonances between the multihole states. Secondly, $F^G_\sigma(m) = \langle X^{pp}_{f_G\sigma}\rangle + \langle X^{qq}_{f_G\sigma}\rangle$ - occupation factors lead to a concentration dependence in both the dispersion, and amplitude of a quasiparticle peak in a spectral density.

From a mathematical point of a view we deal with an equation of the generalized eigenvalues problem, where instead of a customary "matrix of a nonorthogonality" there is a inverse matrix of the relevant occupation factors $F^G_\sigma(m)$. Each $\bar{\alpha}_m$ - rooted vector defines a quasiparticle with a charge and spin 1/2, their local energies are equal to $\Omega^G_m$.

The formula (7) is convenient to calculate the dispersion in a sense that allows obtaining all possible quasiparticle states. However, not all of them can be observed in ARPES- experiment. As is known, the *ARPES*- signal is proportional to the amplitude of a spectral density of a quasiparticle peak:

$$A_\sigma(\vec{k}, E) = \left(-\frac{1}{\pi}\right)\sum_\lambda \text{Im}(G^{\lambda\lambda}_{k\sigma}) = \left(-\frac{1}{\pi}\right)\sum_{\lambda mn} \gamma_{\lambda\sigma}(m)\gamma^+_{\lambda\sigma}(n) \text{Im}(D^{mn}_{k\sigma}(AA) + D^{mn}_{k\sigma}(BB)) \qquad (9)$$

Due to the relevant occupation factors for some quasiparticles the spectral density can be small or even zero. Therefore, in an experiment the relevant quasiparticle peak misses. Because of a large matrix dimension: $\hat{D}_{k\sigma}$ - 32×32, analysis of the spectral density is impossible by an analytical way. A computer simulation of the spectral density was carried out along the basic symmetric directions of the *BZ* at the $T = 0K$. In the *PM* the dispersion and spectral density are obtained with a help of one-sublattice analogs of the formulas (7) and (9).

## III. Quasiparticle spectral density

Fig.2 shows the outcomes of the *ETBM*- computer simulation of a dispersion of the quasiparticle peak on a top of the valence band of $Sr_2CuO_2Cl_2$ and $Ca_2CuO_2Cl_2$ along the basic symmetric directions of the *BZ* in the *AFM* and *PM*- phases at the $T = 0K$. Four terms of the two-hole states determines the composite structure of the valence band. The effects of strong correlations result in turn in unusual properties of this band. The number of states in a band depends on occupation (certainly, by conservation of the number of the all states), the applied field and temperature at the expense of the occupation factors $F_\sigma^G(m)$.

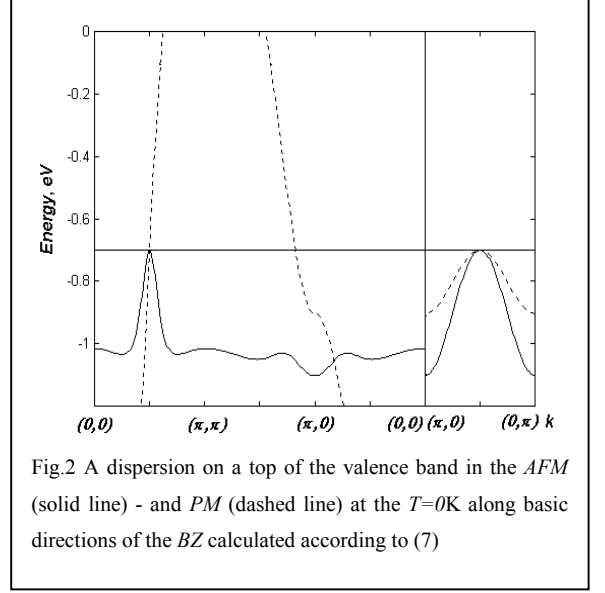

Fig.2 A dispersion on a top of the valence band in the *AFM* (solid line) - and *PM* (dashed line) at the *T=0*K along basic directions of the *BZ* calculated according to (7)

In the *PM* the calculated dispersion (see Fig.2) is similar to one observed in the doped *Bi2212* samples [12]. In the *AFM* the dispersion of the valence band corresponds quantitatively to the ARPES-data for the undoped oxychlorides [13] (see Fig.3) In this respect our approach reproduces existing results for the dispersion of the valence band obtained in a framework of the *t-t'-J*- model [Dagotto]. In addition to the *t-t'-J* - model, a most interesting peculiarity of our dispersion in the *AFM* is a presence of the energy level with a zero spectral density on a top of the valence band. Really, in an undoped antiferromagnetic we deal with $\vec{\alpha}_{0\sigma}(\widetilde{b}_{1\uparrow} \leftrightarrow \widetilde{A}_1)$ - and $\vec{\alpha}_{0\sigma}(\widetilde{b}_{1\downarrow} \leftrightarrow \widetilde{A}_1)$ - quasiparticles on a top of the valence band: At the zero temperature, neglecting of quantum spin fluctuations, the occupation number of the one-hole $|\widetilde{b}_{1\sigma}\rangle$- states for one of projections of a spin

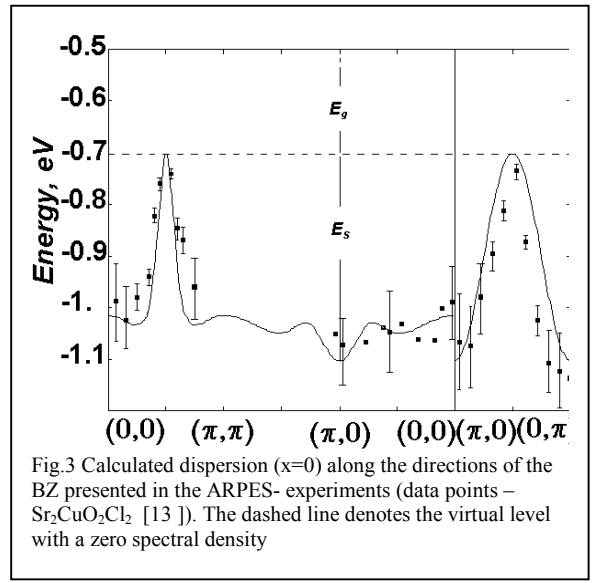

Fig.3 Calculated dispersion (x=0) along the directions of the BZ presented in the ARPES- experiments (data points – $Sr_2CuO_2Cl_2$ [13]). The dashed line denotes the virtual level with a zero spectral density

both in A- , and B-sublattices are equal to zero. Therefore for one of the projections of the spin in the undoped case there is the level with zero spectral weight called in [9] a virtual level. The quasiparticle peak, corresponding to this quasiparticle, is not observed in *ARPES* spectra for the oxychlorides. It is a typical effect of strong correlations. The influence of spin fluctuations will be considered below in Sec.V.

Thus, the energy gap between the valence band and conduction band in the oxychlorides at the $T = 0K$ is possible to present as $E_g(\vec{k}) = E_{ct}(\vec{k}) + E_S(\vec{k})$, where $E_{ct}(\vec{k})$ - a charge transfer gap and $E_S(\vec{k})$ - an energy gap between the virtual level and top of valence band. The gap $E_S(\vec{k})$ has a magnetic nature and is absent in the *PM*. Further for $E_S(\vec{k})$ we shall use the term: "pseudogap". Fig.4 represents a dispersion of $E_S(\vec{k})$ - pseudogap along a boundary of the antiferromagnetic BZ on a background of $|cos(k_x a)-cos(k_y a)|$ - d-like dependence, as it do to show a explicit connection of the pseudogap in antiferromagnetic insulators to the pseudogap in the underdoped *Bi2212* samples and the superconducting gap in the optimal doped *Bi2212* samples [4]. The proximity of the calculated

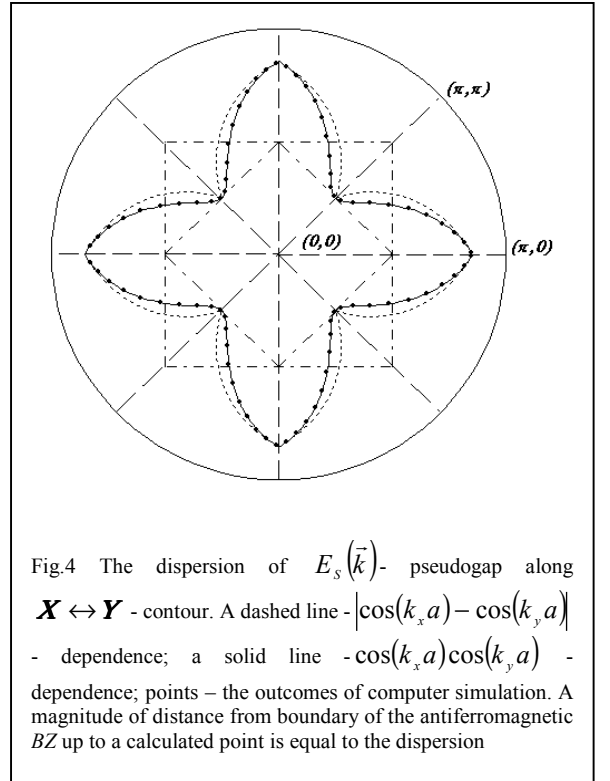

Fig.4 The dispersion of $E_S(\vec{k})$- pseudogap along $X \leftrightarrow Y$ - contour. A dashed line - $|\cos(k_x a) - \cos(k_y a)|$ - dependence; a solid line - $\cos(k_x a)\cos(k_y a)$ - dependence; points – the outcomes of computer simulation. A magnitude of distance from boundary of the antiferromagnetic BZ up to a calculated point is equal to the dispersion

dispersion and observed one allows us to assume, that $\vec{k}$ -contour of "remnant Fermi surface", observed by the authors [2] in $Ca_2CuO_2Cl_2$, can be a simple mark in *ARPES*-spectra from the valence band. A reason for this strange fact can be a pure *2D*-character of quasiparticle states along $\vec{k}$ - contour close to $X \leftrightarrow Y$ - contour. Along all remaining symmetric directions of the *BZ* the calculations reproduce the nonzero contributions to a spectral density from the outplane $d_z$ and $p_z$ - orbitals. The $T_{\lambda\lambda'}(k) = \frac{1}{N}\sum_{R_i} T_{\lambda\lambda'}(\vec{R}_i) e^{ikR_i}$ - factors in (8) do not contain the intersublattice items for $\vec{k}$ along $X \leftrightarrow Y$ -contour and a dispersion can be only ~ $\cos(k_x a)\cos(k_y a)$. In Fig.3 this dependence, as expected, lies down along results of the computer simulation. The square-law dependence near $\overline{M}$ is actually observed in [2], and we suppose, that it is $\cos(k_x a)\cos(k_y a)$ - dependence.

According to our calculations, the reason for the similarity of dispersions in the *AFM* - phase along $\Gamma \leftrightarrow M$ and $X \leftrightarrow Y$ symmetric directions is due to a strong mixing of the valence band state with the triplet $^3B_{1g}$ – state of the underlying band states in the points $\Gamma$ and $M$. The similarity obtained here is determined only by magnitudes of the parameters (distance to $CuO_2$ - plane) concerning with the apical ions Cl or O. In our results there is a strong anisotropy of effective masses near $\vec{k} = \overline{M}$ with $m_{eff}^{XY}/m_{eff}^{\Gamma M} \sim 10$, therefore the similarity concerns only a type of the dispersions along these directions.

A spectral density was calculated along four basic symmetric directions of the *BZ*: $\Gamma \leftrightarrow M$, $M \leftrightarrow X$, $X \leftrightarrow \Gamma$, $X \leftrightarrow Y$ in the *PM* and *AFM*. As follows from (9), a spectral density is additive, therefore there is a possibility to calculate the partial contributions to a spectral density from all orbitals accepted in our calculation:

$$A(\vec{k}, E) = \sum_{\lambda\sigma} A_{\lambda\sigma}(\vec{k}, E),$$

where $\lambda = d_x, b, a, d_z, p_z$. The additive representation is convenient in analysis of polarization *ARPES*-spectra. Fig.5(a,b) represents a $\vec{k}$-dependence of an amplitude of the quasiparticle peak in a spectral density and its partial contributions along $\Gamma \leftrightarrow M$ - symmetric direction in the *AFM* and *PM* accordingly. At the parameters accepted in the calculation [9], the triplet level $^3B_{1g}$ was found on 0.7 eV above energy of a level $A_{1g}$. As was expected, in the *PM* $\vec{k}$-dependence of amplitude of the quasiparticle peak loses a symmetry in a relation to $\overline{M}$ - point. If in $\vec{k} = \Gamma$ the large contribution from the outplane

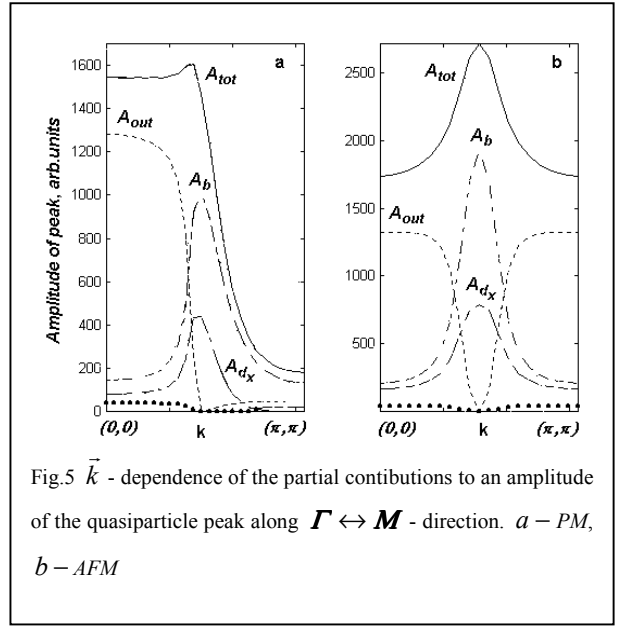

Fig.5 $\vec{k}$ - dependence of the partial contibutions to an amplitude of the quasiparticle peak along $\Gamma \leftrightarrow M$ - direction. $a$ – *PM*, $b$ – *AFM*

orbitals still takes place, an amplitude of the quasiparticle peak in *M* point strongly decreases. The remnant spectral density in this symmetric point is stipulated by submixing of the planar conduction band states: $d_x$ и $b$.

Fig.6(a,b) shows $\vec{k}$-dependence of a amplitude of the quasiparticle peak along the $M \leftrightarrow X$ symmetric direction. In the *AFM*, and *PM* in $\vec{k} = X$ a contribution only from the planar $b$- и $d_x$ -orbitals takes places $A_{pl}(\vec{k}, E) = A_{d_x}(\vec{k}, E) + A_b(\vec{k}, E)$. Contributions from the $p_z$ and $d_z$ -states in this symmetric point

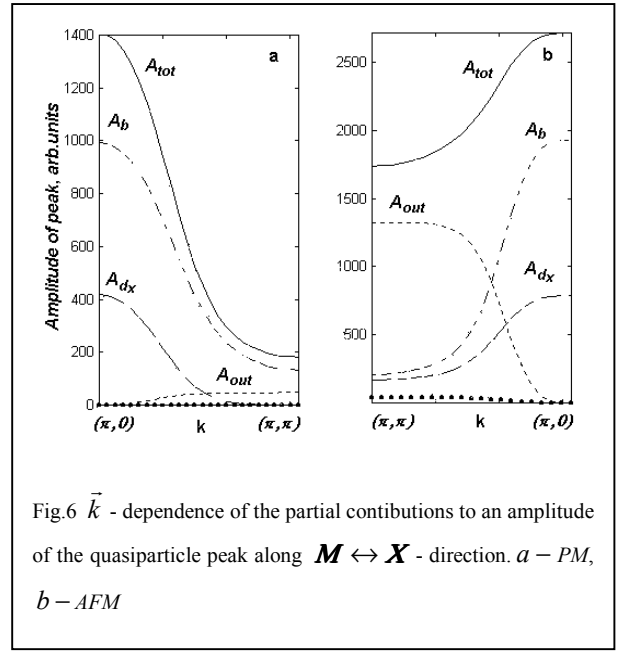

Fig.6 $\vec{k}$ - dependence of the partial contibutions to an amplitude of the quasiparticle peak along $M \leftrightarrow X$ - direction. $a$ – *PM*, $b$ – *AFM*

are absent. The amplitude of the quasiparticle peak monotonically increases along $M \leftrightarrow X$, both in the *AFM,* and *PM*.

A special interest represents $\Gamma \leftrightarrow X$ - symmetric direction (Fig.7(a,b)), where alongside with the same inhibition of the $A_{out}(\vec{k},E)$- partial contribution in $\vec{k}=X$ there is a crossover from the monotonic growth of an amplitude of the quasiparticle peak in a *AFM*- phase to a nonmonotonic behavior with a maximum near $\vec{k}=(2\pi/3,0)$ in the *PM*. The same maximum is observed in *ARPES*-spectra of $Ca_2CuO_2Cl_2$ at the $T=150K$ [4]. Analysis of eigenstates along this symmetric direction has shown, that in the *PM* the maximum in $\vec{k} \sim (2\pi/3,0)$ is caused by an extreme submixing of conduction band states with the same $\vec{k}$. Along $X \leftrightarrow Y$ - symmetric direction both in the *PM,* and *AFM* $\vec{k}$ - dependence of an amplitude of the quasiparticle peak is negligible. Here an amplitude consists of only the $A_{d_x}(\vec{k},E)$- and $A_b(\vec{k},E)$- contributions.

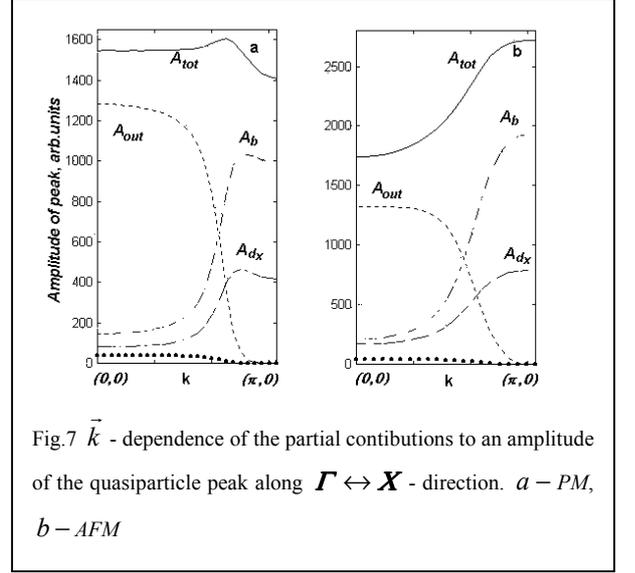

Fig.7 $\vec{k}$ - dependence of the partial contibutions to an amplitude of the quasiparticle peak along $\Gamma \leftrightarrow X$ - direction. $a - PM$, $b - AFM$

Fig.8 and Fig.9 show the spectral density along $\Gamma \leftrightarrow M$ direction in a three-dimensional aspect. There isn't any quasiparticle peak in the region of energies of a virtual level. A contribution from the triplet quasiparticle peak is observed also in the PM.

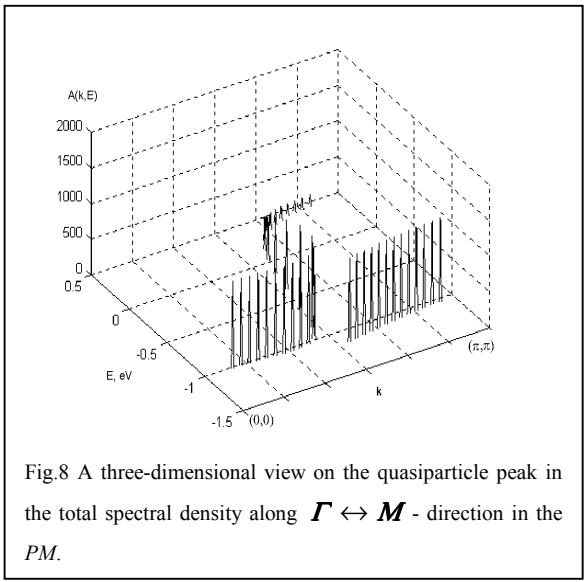

Fig.8 A three-dimensional view on the quasiparticle peak in the total spectral density along $\Gamma \leftrightarrow M$ - direction in the *PM*.

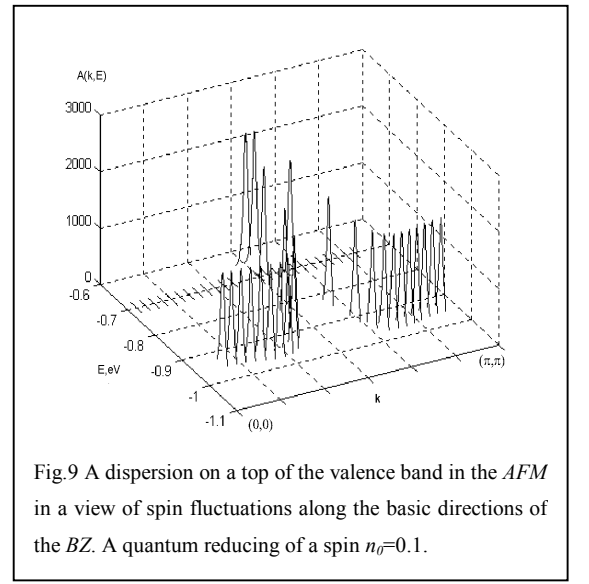

Fig.9 A dispersion on a top of the valence band in the *AFM* in a view of spin fluctuations along the basic directions of the *BZ*. A quantum reducing of a spin $n_0=0.1$.

## IV. Analysis of parity in polarization spectra.

Additional information on a nature of the valence band states can be obtained by analysis of the polarized *ARPES*-spectra. Further we shall analyze polarization dependence of the spectral density (9) in a view of a parity and sizes of the partial contributions. We will neglect effects of a magnetic scattering, as they are less then a charge ones by a factor ($\hbar\omega/mc^2$). However, the charge effects in turn strongly depend on a presence of a magnetic ordering.

Table 1

| | $\Gamma(D_{4h})$ ↔ | $M(D_{4h})$ | $\Gamma$ ↔ | $X(D_{2h})$ |
|---|---|---|---|---|
| $A_{d_x}$ | $odd, (B_{1g})$ | $odd, (B_{1g})$ | $even$ | $even, (A_g)$ |
| $A_b$ | $0, (E_u)$ | $odd, (B_{1g})$ | $0$ | $even, (A_g)$ |
| $A_{in} = A_{d_x} + A_b$ | $0 + odd$ | $odd$ | $0 + even$ | $even$ |
| $A_a$ | $0, (E_u)$ | $even, (A_{1g})$ | $0$ | $odd, (B_{2u})$ |
| $A_{out} = A_{d_z} + A_{p_z} + A_a$ | $\sim even, (A_{1g})$ | $even, (A_{1g})$ | $\sim even$ | $\sim even, (A_g)$ |
| $A_{tot} = A_{in} + A_{out}$ | $\sim even$ | $\sim even$ | $\sim even$ | $\sim even$ |
| possibility of experimental observation: | $\begin{cases} A_{d_x} \\ A_b \text{ seen in the} \\ A_{out} \end{cases}$ $\begin{cases} perp. \\ any \text{ polarization} \\ paral \end{cases}$ | $\begin{cases} A_{in} \\ A_{out} \end{cases}$ seen in the $\begin{cases} perp. \\ paral. \end{cases}$ polarization | $\begin{cases} A_{d_x} \\ A_b \text{ seen in the} \\ A_{out} \end{cases}$ $\begin{cases} paral \\ any \text{ polarization} \\ paral \end{cases}$ | $A_{in}$ and $A_{out}$ seen in the *paral.* polarization |

However, the charge effects in turn strongly depend on a presence of a magnetic ordering.

The table 1 represents $\vec{k}$-groups, small irreducible representations and relevant parities of the plane and outplane partial contributions to the total spectral density in $\Gamma, M, X$-symmetric points in the AFM. The nonzero photocurrent in $\Gamma$-point corresponds to a small $E_u$-irreducible representation of $\vec{k}$-group $D_{4h}$ for $\vec{\alpha}_{0\sigma}(\tilde{b}_{1,\sigma'} \leftrightarrow {}^1\tilde{A}_1)$- quasiparticles, but not to an additional quasiparticle state. A peculiarity of a photocurrent in the center of the *BZ* is its proportionality to the contribution to the spectral density only from the plane oxygen orbitals. The table 1 shows the parity of the total contribution in a spectral density with regard for a size of the partial contributions in the corresponding symmetry points of the *BZ*. The last line in the Table 1 gives indications on possible experimental observation of corresponding partial contributions.

## *V. Effect of spin fluctuations.*

In our nonself-consistent approach a dependence of the electronic structure on the magnetic ordering arises through $F_\sigma(m)$ - occupation factors in the equation (7). In the cause of the unfilled two-hole terms: $F_{\sigma'}(\vec{\alpha}_{i\sigma}) = \langle n_{\tilde{b},\sigma'} \rangle$, so $\vec{\alpha}_{0\sigma}(\tilde{b}_{1,\sigma'} \leftrightarrow {}^1\tilde{A}_1)$- , $\vec{\alpha}_{1\sigma}(\tilde{b}_{1,\sigma'} \leftrightarrow {}^3\tilde{B}_{1,0})$- , $\vec{\alpha}_{2\sigma}(\tilde{b}_{1\sigma} \leftrightarrow {}^3\tilde{B}_{1,2\sigma})$- quasiparticles play the defining role on a top of the valence band in undoped oxichlorides. For nonzero matrix elements $\gamma_{\lambda\sigma}(m)$ (m=0,1.. 31) in the *AFM*, $\langle S_A^z \rangle = \langle S^z \rangle$ in the *A*- sublattice and in the B: $\langle S_B^z \rangle = -\langle S^z \rangle$. The occupation factors can be rewritten as follows: $F_{-\sigma}^G(\vec{\alpha}_{0\sigma}) = F_{-\sigma}^G(\vec{\alpha}_{1\sigma}) = 1/2 - 2\sigma\langle S_G^z \rangle$, $F_\sigma^G(\vec{\alpha}_{2\sigma}) = 1/2 + 2\sigma\langle S_G^z \rangle$, $(\sigma = \pm 1/2)$ where G=A,B. Up to the present moment we were restricted by an Ising consideration, supposing $\langle S^z \rangle = 1/2$. An elementary way to a self-consistent consideration consists in a consruction of an effective Hamiltonian, which will look like a Heisenberg Hamiltonian with the antiferromagnetic interaction, and then to carry out a self-consistent calculation. Thus it is possible to take into account local spin fluctuations (zero fluctuations in the *AFM*), however, to account for nonlocal fluctuations $\langle S_i^+ S_j^- \rangle$ it is required to go beyond a scope of the Hubbard 1 approximation, accepted here. Detailed analysis of applicability of such approach and matching with known results on the magnetic polaron in the $t-t'-J$ model is given in the work [15].

At the expense of the zero spin fluctuations there is, as known, a quantum reducing of the spin: $\langle S^z \rangle = 1/2 - n_0$. A magnitude of $n_0$ can be calculated by the different methods, and in the theory of spin waves for the two-dimensional antiferromagnetic $n_0 \approx 0.2$ [16]. The spin fluctuations essentially change the band structure (see Fig.10). Really, the virtual level has the zero dispersion and zero spectral weight in a framework of the Ising consideration. Due to of the zero spin fluctuations, $F_\downarrow^A(\vec{\alpha}_{0\uparrow}) = n_0$ , $F_\downarrow^B(\vec{\alpha}_{0\uparrow}) = 1 - n_0$ that gives a nonzero dispersion and spectral weight $\sim n_0$.

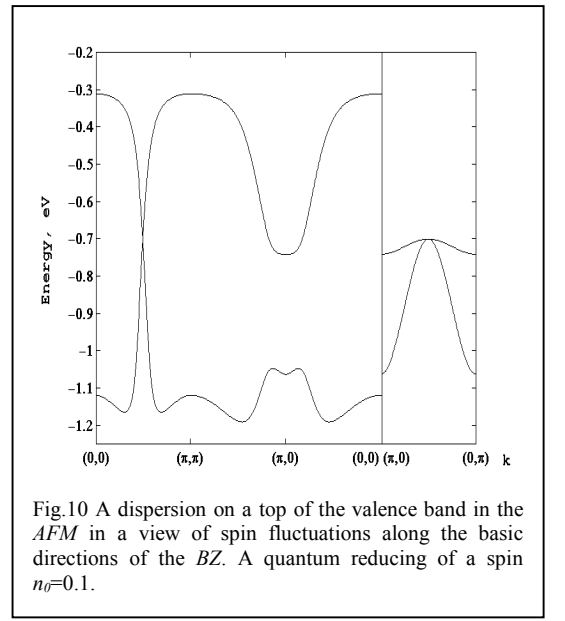

Fig.10 A dispersion on a top of the valence band in the *AFM* in a view of spin fluctuations along the basic directions of the *BZ*. A quantum reducing of a spin $n_0$=0.1.

Thus, the band structure of the undoped *AFM* with spin fluctuations is similar to the band structure the doped Ising antiferromagnetic with the hole concentration $x = n_0$. Because of the small spectral weight the impurity-like subband should be perceived as a low-energy satellite in *ARPES*- spectra. It is possible, that the effects of damping do not allow observing it. Moreover, similar satellites, which intensity increases grows with the temperature are obtained recently by the quantum Monte-Carlo method in the Hubbard model [17]

### *VI. Conclusion*

1. Due to the small energy gap 0.7 eV between the level of the $^3B_{1g}$ - triplet and level of the ZR-singlet in the two-hole sector of Hilbert space, in $\Gamma-$ and $M$ - symmetric points there is a strong hybridization of the singlet and triplet states. The hybridization causes the similarity of the dispersions along $\Gamma \leftrightarrow M$ and $X \leftrightarrow Y$ directions in the AFM. However, our results predict an anisotropy of effective masses in $\vec{k} = \overline{M}$ with a relation $m_{eff}^{XY} / m_{eff}^{\Gamma M} \approx 10$. In the whole a top of the valence band has not a two-dimensional character. As a consequence, at a deviation of an angle of incidence of a parallel-polarized radiation from normal, amplitude of the quasiparticle peak in $\Gamma$ - and $M$ - symmetric points will increase in the *AFM*.

2. On the top of the valence band of oxychlorides: $Sr_2CuO_2Cl_2$ and $Ca_2CuO_2Cl_2$ in the *AFM* there is the pseudogap of magnetic nature: $E_s(\vec{k})\sim 0 \div 0.4$ eV between the virtual level and valence band. The pseudogap converts into a zero in $\overline{M}$. The virtual level has the small spectral density, proportional to a concentration of the zero spin fluctuations $n_0$. Because the dispersion of the pseudogap along $X \leftrightarrow Y$ direction is in the good consent with a dispersion along $\vec{k}$ -contour of the "remnant Fermi surface", we think the one observed in [2] is $\vec{k}$ -contour having only the two-dimension (2D-) dispersion. A contribution to the total spectral density along this direction arrives extremely from the planar $d_x, b$ - orbitals. The pseudogap misses in the *PM*, where the dispersion of the valence band is similar to the dispersion of the optimally doped *Bi2212* [12].

3. A calculated parity of *ARPES*-spectra in $\Gamma, M, X$ points in the AFM according to relative sizes of the partial contributions is even as well as in *ARPES*- experiment [6]. The nonzero photocurrent in $\Gamma$ is due to the small $E_u$ - irreducible representation for $\vec{\alpha}_{0\sigma}(\widetilde{b}_{1,\sigma'} \leftrightarrow {}^1\widetilde{A}_1)$ - quasiparticle in $\vec{k} = \Gamma$, but is not due to an additional satellite state. The peculiarity of a photocurrent in the center of the *BZ* is also its proportionality to a contribution only of the planar oxygen orbitals to the total spectral density.

4. The spin fluctuations essentially change the band structure. Really, the virtual level in the *AFM* has zero dispersion and zero spectral weight in the framework of the Ising consideration only. Thus, the band structure of the oxychlorides due to of spin fluctuations is similar to a band structure of the doped Ising antiferromagnetic. The nonzero spectral weight of a virtual level obtained due to spin fluctuations, results in a weak low-energy satellite in *ARPES*- spectra. Probably effects of damping do not allow observing it on a background of the main peak.

The authors thank A.Lichtenstein and I.Sandalov for useful remarks, D.M.Edwards and A.Oles for attention to the present work, W.Nolting, Tilmann Hickel and Peter Sinjukow for intensive discussions. The work is carried out at support of FCP "Integration" Grant A0019 and RFBR Grant 00-02-16110.